\documentclass[preprint,12pt]{elsarticle}

\usepackage{graphics}
\usepackage{amssymb}
\journal{Comptes Rendus Physique}

\begin{document}

\begin{frontmatter}



\title{Feedback-assisted ponderomotive squeezing}


\author{D. Vitali and P. Tombesi}

\address{School of Science and Technology, Physics Division, \\
University of Camerino, Camerino, and INFN, Sezione di Perugia, Italy, }

\begin{abstract}
We analyze how the radiation pressure interaction between a mechanical element and an intensely driven optical cavity mode can be
exploited for generating squeezed light. We study in particular how the performance of the optomechanical device can be improved when a homodyne-based feedback loop is added to control the motion of the mechanical element of the system. We show that, when driving the cavity at resonance, an appropriate proportional feedback control is able to improve the generation of ponderomotive squeezing, which should be detectable with state-of-the-art apparatuses.

\end{abstract}

\end{frontmatter}

% \linenumbers

\section{Introduction}
\label{sec:intro}

Optomechanical sensors are extensively used for detecting small displacements, masses and forces \cite{vahala-opt-expr,marqPhys,amo09}. The recent progress in nanofabrication techniques now enables these devices to enter a regime where they can manifest quantum effects. As a consequence, a new and highly active research field has emerged with the aim of designing and implementing devices in which a strong and tunable optomechanical interaction allows to manipulate at the quantum level the state of micromechanical resonators and optical modes. Various quantum effects have been predicted \cite{amo09}, and current efforts focus on optically cooling the resonator to its quantum ground state \cite{tobias09,markus09},
and to the more challenging goal of generating entanglement between optical and mechanical modes \cite{epl-pinard,prl07,grob-nature09}.

However, the first prediction of manifestly quantum effects in cavity optomechanical system was made more than fifteen years ago \cite{manc-tomb,fabre} and concerned ponderomotive squeezing, i.e., the possibility to generate quadrature-squeezed light at the cavity output due to the radiation pressure interaction of the cavity mode with a vibrating resonator. In fact, the mechanical element of the cavity is shifted proportionally to the intracavity intensity, and consequently the optical path inside the cavity depends upon such intensity. Therefore the optomechanical system is equivalent to a cavity filled with a nonlinear Kerr medium, which is able to produce squeezed light at the output \cite{milwal}. This analogy between the ponderomotive interaction and Kerr nonlinearity was first pointed out in \cite{dorsel,meystre2} and used to demonstrate radiation-pressure-induced optical bistability \cite{dorsel}. In an ideal Kerr medium the response to the optical field is instantaneous and therefore significative squeezing is produced with a bandwidth limited only by the cavity and with no excess noise. In an optomechanical device, the dynamics of the mechanical element becomes relevant and significantly modifies the squeezing spectrum; the thermal noise acting on the mechanical element represents a limitation, but significant squeezing can still be obtained if radiation pressure effects predominate over thermal noise, which requires large enough cavity finesse and mechanical quality factor.

The first analysis of Refs.~\cite{manc-tomb,fabre}, based on a Fabry-Perot cavity and a description of the mechanical motion as a single harmonic oscillator, was then later extended to the case of many vibrational modes in \cite{pinard}. The problem has been recently reconsidered in a Michelson interferometer setup in \cite{corbitt06}, and a preliminary experimental study of the possible signatures of ponderomotive squeezing in a Fabry-Perot cavity with a movable end-mirror has been recently carried out in \cite{Marin10}.

The experimental demonstration of ponderomotive squeezing would be relevant both at fundamental and at a practical level. In fact, on one hand it would demonstrate the possibility to manipulate in the quantum regime microresonators with microgram masses, and on the other hand it would represent a new source of squeezed light, alternative to optical parametric amplifiers, which could be used, for example, in spatial \cite{fabre2} and spectroscopic measurements \cite{carri}.  In particular squeezing is useful for improving the sensitivity of gravitational wave (GW) interferometers. As first proposed by Caves \cite{caves}, the signal to shot-noise ratio can be improved without increasing the driving power by injecting squeezed vacuum states of light into the signal output port. Later it was realized that injection of squeezed light in the interferometer can also be used to reduce the overall quantum noise including radiation pressure noise, thereby beating the standard-quantum-limit \cite{unruh,reynaud,matsko}. In order to be useful for GW interferometers, a source of squeezed light must work at sideband frequencies within the audio band ($10$ Hz-$10$ KHz), and should have an adjustable phase of squeezing (i.e., the phase with sub-shot noise variance). Standard parametric amplifier sources squeeze light at higher sideband frequencies and at a fixed phase, but recent works have shown an impressive advance in the adaptation of these sources to GW interferometers. In fact, squeezed states have been demonstrated at audio frequencies \cite{audiosquee}, and were tested on a suspended GW prototype interferometer \cite{goda08}. Furthermore, it has been shown that detuned Fabry-Perot cavities can convert a squeezed vacuum with frequency-independent squeeze quadrature into one with frequency-dependent squeeze quadrature \cite{kimble}. These filters have been shown to be broadly applicable to existing interferometer configurations \cite{kimble,chen} and therefore provide the required tunability of the phase of squeezing.

In this paper we study if and how the generation of ponderomotive squeezing can be improved by adding a suitable feedback control to the cavity optomechanical system. In fact, feedback controls have been already proposed and successfully implemented in these systems. A notable example is cold damping feedback \cite{Mancini98,courty,vitalirapcomm,quiescence02} in which a feedback loop adds a viscous force able to cool the resonator, even to its quantum ground state \cite{genes07}, and which has been experimentally applied in various laboratories \cite{cohadon99,arcizet06b,bouwm,rugar,vinante,bowen10}. Moreover, feedback control has been also proposed for suppressing radiation pressure noise in GW interferometers through a ``quantum locking'' scheme \cite{qlock}, and it is therefore interesting to verify if feedback can also help the generation of ponderomotive squeezing. Our analysis will include all fundamental quantum noise sources and thermal noise, and will neglect ``technical'' noise sources, such as laser and electronic noise, even though they can be not negligible in the audio band. We shall see that a proportional feedback control is able to improve squeezing in an appreciable way.

In Sec.~II we provide a quantum Langevin description of optomechanical systems, while in Sec.~III we show how to add feedback controls to them. In Sec.~IV we determine the spectrum of squeezing of the output light and we determine its general properties. In Sec.~V we specialize to the resonant case and we present some numerical results, while Sec.~VI is for concluding remarks.

\section{Quantum Langevin description of the optomechanical system} \label{sec1}

In a typical cavity optomechanical system one has a cavity mode interacting with a mechanical oscillator with a term which is proportional to
the light intensity. The standard situation corresponds to a Fabry-Perot cavity with one heavy, fixed mirror through which a laser of frequency
$\omega _{0}$ drives a cavity mode, and another light end-mirror of mass $m$ (typically in the micro or nanogram range), which is free to oscillate. The optomechanical interaction is provided by the radiation pressure of the cavity mode on the oscillating light mirror. However the same description applies also to related systems, with different geometries and in which the coupling may be provided by different mechanisms. Notable examples are silica toroidal optical microcavities which are coupled to radial vibrational modes of the supporting structure~\cite{vahala-opt-expr}, or partially transparent SiN membranes oscillating within a standard high-finesse cavity \cite{harris,wilson}. In these two systems the coupling is still provided by radiation pressure, while in other cases the coupling has a different origin, as for example in \cite{anetsberger09,vahala-painter09,hong-prl09} where it stems directly from the dipole gradient force.

For clarity we shall refer from now on to the prototypal situation of an optical Fabry-Perot cavity of length $L$ formed by a rigid massive mirror at one end and a vibrating micromechanical mirror at the opposite end, even though the description can be easily adapted to the other configurations. The laser significantly drives only a single cavity mode with frequency $\omega_{c}$, from which it is detuned by $\Delta_{0}=\omega_{c}-\omega_{0}$. The motion of the micro-mirror can be described by the set of its vibrational normal modes, each with its own resonance frequency $\omega_{j}$ and
damping rate $\gamma_{j}$. The Hamiltonian of the system can be written as
\begin{equation}
H=\hbar\omega_{c}a^{\dagger}a+\sum_{j}\frac{\hbar\omega_{j}}{2}(p_{j}%
^{2}+q_{j}^{2})+H_{int} +i\hbar E(a^{\dagger}e^{-i\omega_{0}t} -ae^{i\omega
_{0}t}), \label{ham}%
\end{equation}
where the cavity field annihilation operator $a$ satisfies the commutation relation $\left[  a,a^{\dag}\right]  =1$, and the mechanical modes
are described by dimensionless position and momentum operators satisfying $\left[  q_{k},p_{j}\right]  =i\delta_{kj}$. Denoting by $\kappa$ the
cavity decay rate, the parameter $E$ is related to the input power $\mathcal{P}%
_{in}$ by $ E =\sqrt{2\mathcal{P}_{in}\kappa/\hbar \omega_{0}}$. Phase and amplitude laser noise could be included by assuming that $E$ is a complex stochastic process with fluctuating modulus and phase. The single cavity mode description is valid in the
adiabatic limit when all the relevant mechanical frequencies $\omega_{j}$ are much smaller than the cavity free spectral range $c/2L$, which is
typically satisfied for short cavities. In this limit the scattering of photons by the mirror motion from the driven mode to other cavity
modes is negligible \cite{law}. The interaction between the cavity mode and the vibrational modes is described by $H_{int}$ and it is due to the
radiation pressure acting on the surface $S$ of the vibrating mirror. One has \cite{pinard}
\begin{equation}
\label{radpress}H_{int}=-\int_{S}d^{2} r \vec{P}(\vec{r})\cdot\vec{u}(\vec {r}),
\end{equation}
where $\vec{P}(\vec{r})$ is the radiation pressure field and
\begin{equation}
\label{displfield}\vec{u}(\vec{r})=\sum_{j}\sqrt{\frac{\hbar}{m_{j}\omega_{j}%
}} q_{j}\vec{u}_{j}(\vec{r})
\end{equation}
is the displacement field of the mirror surface at point $\vec{r}$. This field can be written as a sum over the corresponding (dimensionless)
displacement field of each normal mode, $\vec{u}_{j}(\vec{r})$, which is characterized by an effective mass $m_{j}=\rho\int d^{3}r\left\vert
\vec {u}_{j}(\vec{r})\right\vert ^{2}$ ($\rho$ the mirror mass density). We consider a one-dimensional situation, i.e., we assume that the
driving laser and the cavity are perfectly aligned. In this case, light is sensitive only to mirror surface deformations along the cavity axis,
$u_{x}(\vec{r})$, so that Eq.~(\ref{radpress}) becomes
\begin{equation}
\label{radpress2}H_{int}=-\int_{S}d^{2} r P_{x}(\vec{r}) u_{x}(\vec{r}).
\end{equation}
In general, the radiation pressure due to an optical power $\mathcal{P}$ impinging on a mirror with reflection coefficient $\mathcal{R}$ can be
written as
\begin{equation}
\label{radpress3}P_{x}(\vec{r}) = \frac{2 \mathcal{P}}{c} \mathcal{R} v_{opt}^{2}(\vec{r}),
\end{equation}
with $v_{opt}(\vec{r})$ denoting the spatial structure of the incident optical field on the mirror surface. Within the cavity, one can rewrite
$2 \mathcal{P}/c=\hbar(\omega_{c}/L)a^{\dagger} a $ and also assume $\mathcal{R} \simeq1$. One ends up with
\begin{equation}
\label{radpress4}H_{int}=-\hbar\sum_{j} G_{0}^{j}a^{\dagger}aq_{j},
\end{equation}
where the optomechanical couplings are given by
\begin{equation}
G_{0}^{j}=\frac{\omega_{c}c_{j}}{L} \sqrt{\frac{\hbar}{m_{j}\omega_{j}}},
\label{coupls}%
\end{equation}
and
\begin{equation}
c_{j} = \int_{S}d^{2} r v_{opt}^{2}(\vec{r}) (u_{j})_{x}(\vec{r})
\end{equation}
is the overlap at the mirror surface between the cavity mode and the $j$-th mechanical mode. Due to the chosen normalization of
$v_{opt}^{2}(\vec{r})$ and $\vec{u}_{j}(\vec{r})$, the overlaps satisfy the condition $-1 \leq c_{j} \leq 1$.
Eqs.~(\ref{radpress4})-(\ref{coupls}) show that the radiation pressure directly couples the cavity mode only with the mirror collective
displacement operator $q_{eff}=\sum_{j} G_{0}^{j}q_{j}$. When the detection bandwidth involves only a single, isolated, vibrational normal mode
of the microresonator, the collective coordinate $q_{eff}$ is well approximated by the selected normal mode, and the single harmonic oscillator
description usually adopted is justified. In the most general case, one has to include in the dynamical description of the system all the
vibrational normal modes which contribute to the detected signal.

The unavoidable action of damping and noise onto the dynamics associated with the Hamiltonian of Eq.~(\ref{ham}) is described by adopting the
formalism of quantum Langevin equations \cite{gard,giov} which, in the frame rotating at the laser frequency $\omega_{0}$, are given by
\begin{eqnarray}
\dot{q}_{j}  &  =& \omega_{j}p_{j}, \label{QLEnonlinear1}\\
\dot{p}_{j}  &  = & -\omega_{j}q_{j}-\gamma_{j}p_{j}+G_{0}^{j}a^{\dag}a+\xi
_{j},\\
\dot{a}  &  =& -(\kappa+i\Delta_{0})a+i\sum_{j} G_{0}^{j}aq_{j}+E+\sqrt{2\kappa }a^{in}. \label{QLEnonlinear3}
\end{eqnarray}
The input noise $ a^{in}(t)$ describes the optical vacuum field entering the cavity, and it is delta correlated in the time domain $\langle a^{in}(t)a^{in,\dag}(t^{\prime})\rangle=\delta(t-t^{\prime})$ \cite{gard}, while the
mechanical Brownian stochastic forces with zero mean value $\xi_{j}(t)$ are uncorrelated from each other and have the following, generally
non-Markovian, correlation functions \cite{giov}
\begin{equation}\label{browncorre}
\langle\xi_{k}(t)\xi_{j}(t^{\prime})\rangle=\delta_{kj}\frac{\gamma_{j}}%
{2\pi\omega_{j}}\int d\omega e^{-i\omega(t-t^{\prime})}\omega\left[ \coth\left(  \frac{\hbar\omega}{2k_{B}T}\right)  +1\right]  ,
\end{equation}
with $k_{B}$ the Boltzmann constant and $T$ is the temperature of the reservoir of the micromechanical mirror.

We are interested in ponderomotive squeezing and therefore one requires a strong radiation pressure interaction, which is achieved when the
intracavity field is very intense, i.e., for high-finesse cavities and enough driving power. In this limit (and if the system is stable) the
system is characterized by a semiclassical steady state with the cavity mode in a coherent state with amplitude $\alpha_{s}$ ($|\alpha_{s}|
\gg1$), and a new equilibrium position for the vibrational modes, displaced by $q_{s}^{j}$. The parameters $\alpha_{s}$ and $q_{s}^{j}$ are the
solutions of the nonlinear algebraic equations obtained by factorizing Eqs.~(\ref{QLEnonlinear1})-(\ref{QLEnonlinear3}) and setting the time
derivatives to zero. They are given by
\begin{eqnarray}
 q_{s}^{j}& = & \frac{G_{0}^{j}|\alpha_{s}|^{2}}{\omega_{j}},\\
  p_{s}^{j}& = & 0,\\
\alpha_{s}& = & \frac{E}{\kappa+i\Delta},
\end{eqnarray}
where the effective detuning $\Delta$ is obtained from $\Delta_{0}$ by subtracting the frequency shift caused by the steady state
radiation pressure
\label{delta}%
\begin{equation}
\Delta=\Delta_{0}-|\alpha_{s}|^{2}\sum_{j}\frac{[G_{0}^{j}]^{2}}%
{\omega_{j}}.
\end{equation}
These steady state equations are responsible for the optical bistability observed in \cite{dorsel} and analyzed in \cite{meystre2}.
Then, we linearize Eqs. (\ref{QLEnonlinear1})-(\ref{QLEnonlinear3}) around the steady state values by writing operators as sums of averages plus
fluctuations: $a=\alpha_{s}+\delta a$, $q_{j}=q_{s}^{j}+\delta q_{j}$ and $p_{j}=p_{s}^{j}+\delta p_{j}$. The nonlinear terms $\delta
a^{\dag}\delta a$ and $\delta a\delta q_{j}$ can be ignored when the fluctuations are much smaller than the mean value, and this is certainly
satisfied when $|\alpha _{s}| \gg1$. One therefore arrives at a system of linearized quantum Langevin equations
\begin{eqnarray}
\delta\dot{q}_{j}  &  = &\omega_{j}\delta p_{j}, \label{QLElinear1}\\
\delta\dot{p}_{j}  &  = & -\omega_{j}\delta q_{j}-\gamma_{j}\delta p_{j}%
+G_{j}\delta X_a+\xi_{j}, \label{QLElinear2}\\
\delta\dot{X}_a  &  = & -\kappa\delta X_a+\Delta \delta Y_a+\sqrt{2\kappa}%
X_a^{in},\\
\delta\dot{Y}_a  &  = & -\kappa\delta Y_a-\Delta \delta X_a+\sum_j G_{j}\delta q_{j}+\sqrt{2\kappa}Y_a^{in}. \label{QLElinear4}
\end{eqnarray}
We have chosen the phase reference of the cavity field so that $\alpha_{s}$ is real and positive, we have defined the cavity field quadratures $\delta
X_a\equiv(\delta a+\delta a^{\dag})/\sqrt{2}$ and $\delta Y_a\equiv(\delta a-\delta a^{\dag})/i\sqrt{2}$ and the corresponding Hermitian input noise
quadratures $X_a^{in}\equiv(a^{in}+a^{in,\dag})/\sqrt{2}$ and $Y_a^{in}%
\equiv(a^{in}-a^{in,\dag})/i\sqrt{2}$. We have also defined the effective optomechanical couplings
\begin{equation}
G_{j}\equiv G_{0}^{j} \alpha_{s}\sqrt{2}=\frac{2\omega_{c}c_j}{L}
\sqrt{\frac{\mathcal{P}_{in} \kappa}{m_{j} \omega_{j} \omega_{0} \left(
\kappa^{2}+\Delta^{2}\right)  }}. \label{optoc}
\end{equation}

\section{Adding the feedback loop}

The fluctuation dynamics described by Eqs.~(\ref{QLElinear1})-(\ref{QLElinear4}) is then modified by adding a feedback force acting on the vibrational modes of the resonator (see Fig.~1), that can be applied in various ways, either exploiting the radiation pressure force of an additional laser beam (as in Ref.~\cite{cohadon99}), or through electromechanical actuators as in \cite{bouwm,rugar,vinante,bowen10}. The feedback loop is obtained by extracting a fraction of the cavity output which is then processed in order to drive an appropriate actuator acting on the resonator. The simplest and most efficient way to extract the feedback loop mode is by means of a beam splitter with amplitude transmission $t$ and reflection $r$, with $t^2+r^2=1$ (we can always choose mode phases so that $t$ and $r$ are real), so that
\begin{eqnarray}
  d &=& t a_{out}-r b_{in} \label{dmode},\\
  c &=& r a_{out}+t b_{in}. \label{cmode}
\end{eqnarray}
$d$ is the annihilation operator describing the overall output of the device, and $c$ is the annihilation operator of the light employed for the feedback loop. This latter optical mode is detected by a balanced homodyne detector measuring with quantum efficiency $\eta$ the quadrature with phase $\theta$. Detection with non-unit efficiency $\eta$ can always be described in terms of an effective beam splitter with amplitude transmission $\sqrt{\eta}$ in front of a perfect detector \cite{gard}, so that the detected mode is given by
\begin{equation}\label{smode}
    s=\sqrt{\eta} c+\sqrt{1-\eta} v_{in},
\end{equation}
where $v_{in}$ is the annihilation operator describing the optical vacuum noise unavoidably entering the detector. In practice the bosonic mode $s$ describes the photocurrent at the output of the homodyne detector which is then electronically filtered and amplified in order to actuate the vibrational modes (see Fig.~1).

\begin{figure}[tb]
\centerline{\includegraphics[width=0.90\textwidth]{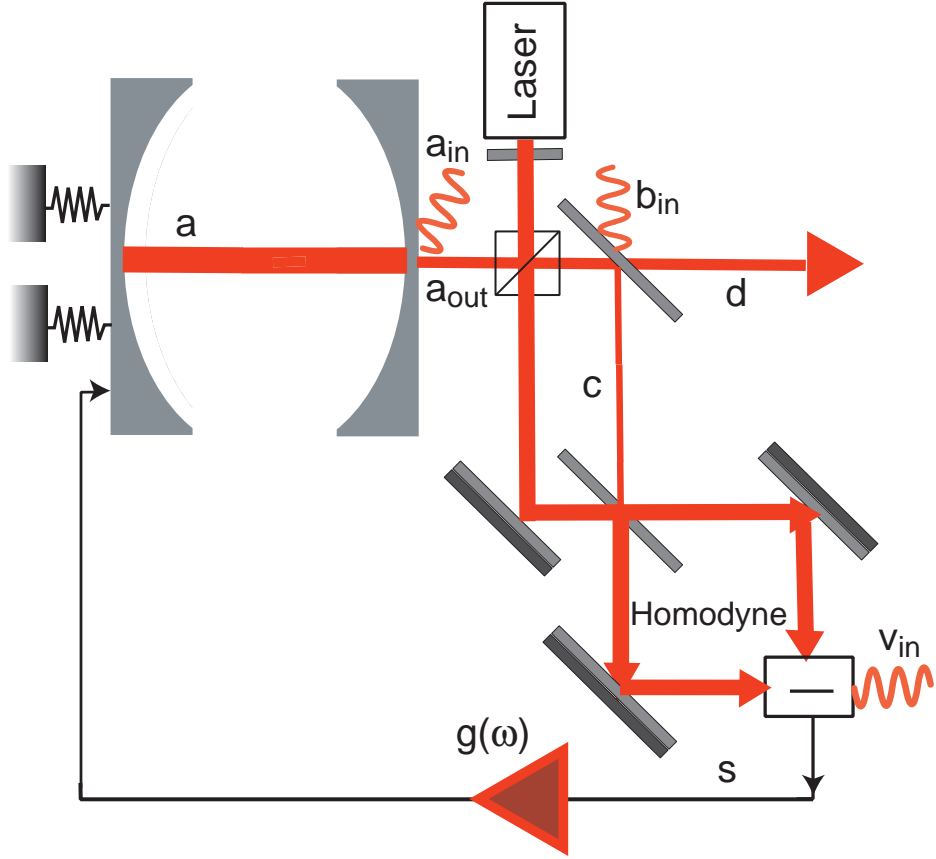}} \caption{Schematic description of a cavity optomechanical system subject to a homodyne-mediated feedback control of the mechanical resonator.}\label{fig1}
\end{figure}

The feedback loop generally acts differently on the different mechanical modes and it is described by an additional force term on Eq.~(\ref{QLElinear2}), given by the time convolution
\begin{equation}\label{feedforcej}
  \left\{\delta\dot{p}_{j}(t)\right\}_{fb}=- \int_{-\infty}^t dt' g_j(t-t') \frac{\theta_s(t')}{\sqrt{2\kappa}},
\end{equation}
where $g_j(t)$ is the causal feedback transfer function on the $j$-th vibrational normal mode and $\theta_s(t)= (s e^{-i \theta}+s^{\dagger} e^{i \theta})/\sqrt{2}$ is the detected field quadrature. The homodyne phase $\theta$ and the transfer functions $g_j$ are the feedback loop parameters which must be optimized in order to achieve the best possible squeezing of the output mode $d$. From Eqs.~(\ref{dmode})-(\ref{smode}) and using the input-output relation $a_{out}(t)=\sqrt{2 \kappa} a(t)-a_{in}(t)$ \cite{gard}, one has
\begin{equation}\label{smodefin}
    s(t)=\sqrt{2\kappa \eta} r a(t)-\sqrt{\eta} r a_{in}(t)+\sqrt{\eta} t b_{in}(t)+\sqrt{1-\eta} v_{in}(t).
\end{equation}
Inserting Eq.~(\ref{smodefin}) into Eq.~(\ref{feedforcej}) and then adding the resulting feedback force into the quantum Langevin equations of Eqs.~(\ref{QLElinear1})-(\ref{QLElinear4}), we arrive at the following Fourier-transformed equations
\begin{eqnarray}
-i\omega \delta q_{j}(\omega)  &  = &\omega_{j}\delta p_{j}(\omega), \label{QLElinearft1}\\
-i\omega \delta p_{j}(\omega)  &  = & -\omega_{j}\delta q_{j}(\omega)-\gamma_{j}\delta p_{j}(\omega)
+\left[G_{j}-r\sqrt{\eta}g_j(\omega)\cos\theta\right]\delta X_a(\omega) \nonumber\\
&& -  r\sqrt{\eta}g_j(\omega)\sin\theta \delta Y_a(\omega) +\xi_{j}(\omega)+\frac{g_j(\omega)}{\sqrt{2\kappa}}n_{fb}(\omega),\label{QLElinearft2} \\
-i \omega \delta X_a(\omega)  &  = & -\kappa\delta X_a(\omega)+\Delta\delta Y_a(\omega)+\sqrt{2\kappa}X_a^{in}(\omega),\label{QLElinearft3} \\
-i \omega \delta Y_a(\omega)  &  = & -\kappa\delta Y_a(\omega)-\Delta\delta X_a(\omega)+\sum_j G_{j}\delta q_{j}(\omega)+\sqrt{2\kappa}Y_a^{in}(\omega), \label{QLElinearft4}
\end{eqnarray}
where
\begin{equation}\label{feednoise}
    n_{fb}(\omega)=\sqrt{\eta} r \theta_a^{in}(\omega)-\sqrt{\eta} t \theta_b^{in}(\omega)-\sqrt{1-\eta}\theta_v^{in}(\omega)
\end{equation}
is the vacuum input noise injected by the feedback loop, and we have defined the Fourier transform of the field quadratures of a given mode with annihilation operator $f(\omega)$, $\theta_f^{in}(\omega)= [f(\omega) e^{-i \theta}+f^{\dagger}(\omega) e^{i \theta}]/\sqrt{2}$. Notice that, due to the fact that we have written the quantum Langevin equations in the frame rotating at the laser frequency $\omega_0$, the frequency $\omega$ is referred to this latter frequency, i.e., $\omega=0$ corresponds to the laser frequency.

The observables of interest are the Fourier transform of the field quadratures of the output mode $d$ given by Eq.~(\ref{dmode}). The explicit expression of these quadratures as a function of the cavity, mechanical, and feedback loop parameters can be obtained by replacing into Eq.~(\ref{dmode}) the input-output relation and the explicit solution of Eqs.~(\ref{QLElinearft1})-(\ref{QLElinearft4}) for the cavity field amplitude quadrature $\delta X_a(\omega)$ and the phase quadrature $\delta Y_a(\omega)$. After long, but straightforward calculations one gets the Fourier transform of these quadratures as a linear combination of all the noise terms acting on the system
\begin{eqnarray}
  X_d(\omega) &=&  \sigma_1(\omega)X_a^{in}(\omega)+ \sigma_2(\omega)Y_a^{in}(\omega)+ \sigma_3(\omega)X_b^{in}(\omega)+ \sigma_4(\omega)Y_b^{in}(\omega)
  \nonumber \\
  && +\sigma_5(\omega)\theta_v^{in}(\omega)+\sigma_6(\omega)\sum_j G_j\chi_j^{(0)}(\omega)\xi_j(\omega),\label{xdgen}\\
   Y_d(\omega) &=&  \mu_1(\omega)X_a^{in}(\omega)+ \mu_2(\omega)Y_a^{in}(\omega)+ \mu_3(\omega)X_b^{in}(\omega)+ \mu_4(\omega)Y_b^{in}(\omega)
  \nonumber \\
  && +\mu_5(\omega)\theta_v^{in}(\omega)+\mu_6(\omega)\sum_j G_j\chi_j^{(0)}(\omega)\xi_j(\omega)\label{ydgen},
\end{eqnarray}
where the coefficients are given by
\begin{eqnarray}
  \sigma_1(\omega) &=& t D(\omega)^{-1}\left[\left(\kappa+i \omega \right)\left[\kappa-i \omega+r\sqrt{\eta}\sin \theta \lambda_g(\omega)\right]-\Delta\left[\Delta -\lambda_G(\omega)\right]\right],\\
   \sigma_2(\omega) &=& t \Delta D(\omega)^{-1}\left[2\kappa+r\sqrt{\eta}\sin \theta \lambda_g(\omega)\right],\\
  \sigma_3(\omega) &=& -D(\omega)^{-1} \left[\Delta \sqrt{\eta}\cos \theta \lambda_g(\omega) +r\left(\kappa-i \omega \right)\left[\kappa-i \omega+r\sqrt{\eta}\sin \theta \lambda_g(\omega)\right]\right.\nonumber \\
  &&\left.+r\left[\Delta^2
  -\Delta\lambda_G(\omega)\right]\right],\\
  \sigma_4(\omega) &=&  -D(\omega)^{-1}\left[t^2\Delta \sqrt{\eta}\sin \theta \lambda_g(\omega)\right],\\
  \sigma_5(\omega) &=&  -t\Delta D(\omega)^{-1}\sqrt{1-\eta} \lambda_g(\omega),\\
   \sigma_6(\omega) &=&  t\Delta D(\omega)^{-1}\sqrt{2\kappa},
\end{eqnarray}
for the output amplitude quadrature and
\begin{eqnarray}
  \mu_1(\omega) &=& -t D(\omega)^{-1}\left[2\kappa \Delta-2\kappa \lambda_G+(\kappa+i\omega)r\sqrt{\eta}\cos \theta \lambda_g(\omega)\right],\\
   \mu_2(\omega) &=& t D(\omega)^{-1}\left[\kappa^2+ \omega^2 -\Delta\left[\Delta -\lambda_G(\omega)+r\sqrt{\eta}\cos \theta \lambda_g(\omega)\right]\right],\\
   \mu_3(\omega) &=&  -D(\omega)^{-1}t^2 \left[(\kappa-i\omega)\sqrt{\eta}\cos \theta \lambda_g(\omega)\right],\\
   \mu_4(\omega) &=& -D(\omega)^{-1}\left[(\kappa-i\omega)\sqrt{\eta}\sin \theta \lambda_g(\omega)\right.\nonumber \\
  && \left.+r\left[(\kappa-i \omega)^2+\Delta^2
  -\Delta\lambda_G(\omega)+\Delta r\sqrt{\eta}\cos \theta \lambda_g(\omega)\right]\right],\\
  \mu_5(\omega) &=&  -t D(\omega)^{-1}\left[(\kappa-i\omega) \sqrt{1-\eta} \lambda_g(\omega)\right],\\
   \mu_6(\omega) &=&  t D(\omega)^{-1}\left[(\kappa-i\omega) \sqrt{2\kappa}\right],
\end{eqnarray}
for the phase quadrature. We have used the definition
\begin{equation}\label{denom}
    D(\omega)=\left(\kappa-i \omega \right)\left[\kappa-i \omega+r\sqrt{\eta}\sin \theta \lambda_g(\omega)\right]+\Delta\left[\Delta
  -\lambda_G(\omega)+r\sqrt{\eta}\cos \theta \lambda_g(\omega)\right],
\end{equation}
and we have introduced the two frequency-dependent quantities
\begin{eqnarray}
  \lambda_g(\omega) &=& \sum_j G_j\chi_j^{(0)}(\omega)g_j(\omega) \\
  \lambda_G(\omega) &=& \sum_j G_j^2\chi_j^{(0)}(\omega),\label{lamG}
\end{eqnarray}
depending upon the unperturbed susceptibility of the j-th vibrational mode
\begin{equation}\label{susczero}
    \chi_j^{(0)}(\omega)=\frac{\omega_j}{\omega_j^2-\omega^2-i\omega \gamma_j}.
\end{equation}
The function $\lambda_g(\omega)$ describes the effect of feedback, while $\lambda_G(\omega)$  plays the role of an effective mechanical susceptibility of the resonator (see also \cite{pinard}).

\section{The squeezing spectrum}
In order to characterize the present optomechanical device as a source of squeezed light we have to determine the quadrature noise spectrum of the output mode $d$ which, for a generic quadrature $\phi_d(\omega)$ is defined, due to stationarity, by the relation
\begin{equation}\label{spephi}
    \left\langle \phi_d(\omega) \phi_d(\omega')\right\rangle =S_d^{\phi}(\omega)\delta (\omega+\omega').
\end{equation}
The output light is squeezed at phase $\phi$ when $S_d^{\phi}(\omega)<1/2$, i.e., when the noise spectrum is below the shot-noise limit (equal to $1/2$ with our definitions). The quadrature noise spectrum $S_d^{\phi}(\omega)$ can be written in terms of the amplitude and phase quadrature noise spectra, $S_d^{X}(\omega)$ and $S_d^{Y}(\omega)$ respectively, and their correlation spectrum $S_d^{XY}(\omega)$\footnote{$S_d^{XY}(\omega)\delta(\omega+\omega')=\left[\left\langle X_d(\omega)Y_d(\omega')\right\rangle+\left\langle Y_d(\omega')X_d(\omega)\right\rangle +\left\langle X_d(\omega')Y_d(\omega)\right\rangle +\left\langle Y_d(\omega)X_d(\omega')\right\rangle\right]/4$}, as
\begin{equation}\label{spephi2}
    S_d^{\phi}(\omega)= \frac{S_d^{X}(\omega)+S_d^{Y}(\omega)}{2}+\frac{S_d^{X}(\omega)-S_d^{Y}(\omega)}{2}\cos 2\phi+S_d^{XY}(\omega)\sin 2\phi.
\end{equation}
However, rather than looking at the noise spectrum at a fixed phase of the field, one can perform an optimization and consider, for every frequency $\omega$, the field phase $\phi_{opt}(\omega)$ possessing the minimum noise spectrum $S_d^{\phi}(\omega)$: in this way one defines the \emph{optimal squeezing spectrum} which can be easily obtained by minimizing Eq.~(\ref{spephi2}),
\begin{eqnarray}\label{speopt}
    S_d^{opt}(\omega)&=&\min_{\phi}S_d^{\phi}(\omega)=\frac{1}
    {2}\left[S_d^{X}(\omega)+S_d^{Y}(\omega)-\sqrt{\left[S_d^{X}(\omega)-S_d^{Y}(\omega)\right]^2+
    4\left[S_d^{XY}(\omega)\right]^2}\right] \nonumber \\
    &&=\frac{2S_d^{X}(\omega)S_d^{Y}(\omega)-2\left[S_d^{XY}(\omega)\right]^2}{S_d^{X}(\omega)+S_d^{Y}(\omega)+\sqrt{\left[S_d^{X}(\omega)
    -S_d^{Y}(\omega)\right]^2+
    4\left[S_d^{XY}(\omega)\right]^2}}.
\end{eqnarray}
The frequency-dependent optimal phase is correspondingly given by
\begin{equation}\label{phiopt}
   \phi_{opt}(\omega)=\frac{1}{2}\arctan\left[\frac{2S_d^{XY}(\omega)}{S_d^{X}(\omega)-S_d^{Y}(\omega)}\right].
\end{equation}
We have to remember that these noise spectra satisfy the Heisenberg uncertainty theorem, originating from the fact that the two quadratures $X_d$ and $Y_d$ are non commuting observables. This fact constraints their variances and the corresponding noise spectra according to \cite{bragbook}
\begin{equation}\label{heisen}
    S_d^{X}(\omega)S_d^{Y}(\omega)-\left[S_d^{XY}(\omega)\right]^2 > \frac{1}{4}.
\end{equation}
The squeezing properties of the device are fully determined by the three noise spectra $S_d^{X}(\omega)$, $S_d^{Y}(\omega)$, and $S_d^{XY}(\omega)$, which can be obtained from Eqs.~(\ref{xdgen})-(\ref{ydgen}) and the knowledge of the spectrum of all the noises affecting the system. Since the various noise terms are all uncorrelated, one has the following general expressions
\begin{eqnarray}
  && S_d^{X}(\omega) = \left|\sigma_1(\omega)\right|^2 S_{a,in}^{X}(\omega)+ \left|\sigma_2(\omega)\right|^2 S_{a,in}^{Y}(\omega)+ 2{\rm Re}\left\{\sigma_1(\omega)\sigma_2^{*}(\omega)\right\} S_{a,in}^{XY}(\omega)\nonumber \\
  &&+\left|\sigma_3(\omega)\right|^2 S_{b,in}^{X}(\omega)+ \left|\sigma_4(\omega)\right|^2 S_{b,in}^{Y}(\omega)+ 2{\rm Re}\left\{\sigma_3(\omega)\sigma_4^{*}(\omega)\right\} S_{b,in}^{XY}(\omega)\nonumber \\
  &&+\left|\sigma_5(\omega)\right|^2 S_{v,in}^{\theta}(\omega)+\left|\sigma_6(\omega)\right|^2\coth\left(\frac{\hbar \omega}{2\kappa_B T}\right)
  {\rm Im}\left\{\lambda_G(\omega)\right\},\label{spexdgen}\\
  && S_d^{Y}(\omega) = \left|\mu_1(\omega)\right|^2 S_{a,in}^{X}(\omega)+ \left|\mu_2(\omega)\right|^2 S_{a,in}^{Y}(\omega)+ 2{\rm Re}\left\{\mu_1(\omega)\mu_2^{*}(\omega)\right\} S_{a,in}^{XY}(\omega)\nonumber \\
  &&+\left|\mu_3(\omega)\right|^2 S_{b,in}^{X}(\omega)+ \left|\mu_4(\omega)\right|^2 S_{b,in}^{Y}(\omega)+ 2{\rm Re}\left\{\mu_3(\omega)\mu_4^{*}(\omega)\right\} S_{b,in}^{XY}(\omega)\nonumber \\
  &&+\left|\mu_5(\omega)\right|^2 S_{v,in}^{\theta}(\omega)+\left|\mu_6(\omega)\right|^2\coth\left(\frac{\hbar \omega}{2\kappa_B T}\right)
  {\rm Im}\left\{\lambda_G(\omega)\right\},\label{speydgen}\\
  &&S_d^{XY}(\omega) = {\rm Re}\left\{\sigma_1(\omega)\mu_1^{*}(\omega)\right\} S_{a,in}^{X}(\omega)+ {\rm Re}\left\{\sigma_2(\omega)\mu_2^{*}(\omega)\right\} S_{a,in}^{Y}(\omega)\nonumber \\
  &&+ {\rm Re}\left\{\sigma_1(\omega)\mu_2^{*}(\omega)+\sigma_2(\omega)\mu_1^{*}(\omega)\right\} S_{a,in}^{XY}(\omega)+{\rm Re}\left\{\sigma_3(\omega)\mu_3^{*}(\omega)\right\} S_{b,in}^{X}(\omega)\label{spexydgen} \\
  &&+ {\rm Re}\left\{\sigma_4(\omega)\mu_4^{*}(\omega)\right\} S_{b,in}^{Y}(\omega)+ {\rm Re}\left\{\sigma_3(\omega)\mu_4^{*}(\omega)+\sigma_4(\omega)\mu_3^{*}(\omega)\right\}  S_{b,in}^{XY}(\omega) \nonumber\\
  &&+{\rm Re}\left\{\sigma_5(\omega)\mu_5^{*}(\omega)\right\} S_{v,in}^{\theta}(\omega)+{\rm Re}\left\{\sigma_6(\omega)\mu_6^{*}(\omega)\right\}\coth\left(\frac{\hbar \omega}{2\kappa_B T}\right)
  {\rm Im}\left\{\lambda_G(\omega)\right\}.\nonumber
  \end{eqnarray}
Here we have assumed generic spectra for the input noise entering the cavity $a_{in}$ and for the one entering the beam splitter $b_{in}$. We have also employed the thermal Brownian noise correlation function of Eq.~(\ref{browncorre}) and the definitions of Eqs.~(\ref{lamG})-(\ref{susczero}). The fact that the thermal Brownian noise spectrum is proportional to ${\rm Im}\left\{\lambda_G(\omega)\right\}$ is a consequence of the fluctuation-dissipation theorem \cite{Landau}.

These equations give the most general expression for the output spectrum of squeezing for a cavity optomechanical system subject to feedback. The best ponderomotive squeezing is achieved by minimizing $S_d^{opt}(\omega)$ of Eq.~(\ref{speopt}). However, such a minimization is subject to two constraints. The first constraint comes from stability conditions. In fact, we are considering the quantum fluctuations of the optomechanical system around its stationary state, which is possible only if the system parameters are within the stability region. When the system is stable there is no exponential divergence in its time evolution and this is possible only if all the poles of the solutions in the frequency domain of Eqs.~(\ref{xdgen})-(\ref{ydgen}) have negative imaginary part. From the explicit solutions of the coefficients $\sigma_i(\omega)$ and $\mu_i(\omega)$ one easily sees that stability is guaranteed when all the zeros of $D(\omega)$ of Eq.~(\ref{denom}) have negative imaginary part. Eq.~(\ref{denom}) shows that these stability conditions are generally quite involved and strongly depend upon the explicit form of $\lambda_g(\omega)$, i.e., of the feedback transfer functions $g_j(\omega)$. The second constraint comes from the fact that all the feedback transfer functions $g_j(t)$ must be causal, as it is evident in Eq.~(\ref{feedforcej}).

\section{Ponderomotive squeezing at resonance}

Let us now apply the general description of the preceding sections to the case when the driving laser is locked at resonance with the cavity mode, i.e., when $\Delta=0$. First of all we consider the standard situation of input optical noises $a_{in}$ and $b_{in}$ and $v_{in}$ in the vacuum state, implying
\begin{eqnarray}\label{vacuuminput}
    S_{a,in}^{X}(\omega) &=& S_{a,in}^{Y}(\omega)=S_{b,in}^{X}(\omega)=S_{b,in}^{Y}(\omega)=S_{v,in}^{\theta}(\omega)=\frac{1}{2}, \\
    S_{a,in}^{XY}(\omega) &=& S_{b,in}^{XY}(\omega)=0.
\end{eqnarray}
The resonant case is particularly convenient not only because the description considerably simplifies, but especially because the limitations imposed by the stability conditions are much less stringent. In fact in this case the relevant zeros are the solution of
\begin{equation}\label{stabres}
\kappa-i \omega+r\sqrt{\eta}\sin \theta \lambda_g(\omega)=0.
\end{equation}
The sign of the imaginary part of the solutions depends upon the explicit form of $\lambda_g(\omega)$, but it can be verified that stability is much easier verified with respect to the off-resonant case. In particular the system is always stable without feedback, $r \lambda_g(\omega)=0$, or when $\sin\theta=0$.

The most relevant simplification of the resonant case is that the amplitude fluctuations $\delta X$ of the cavity field are decoupled from the other variables (i.e., both from phase fluctuations and from the resonator motion); as a consequence one has simply
\begin{equation}\label{speampli}
    S_{d}^{X}(\omega)=\frac{1}{2},
\end{equation}
as it can be verified from Eq.~(\ref{spexdgen}) and the fact that when $\Delta=0$, $\sigma_2(\omega)=\sigma_4(\omega)=\sigma_5(\omega)=\sigma_6(\omega)=0$, $\sigma_1(\omega)=t(\kappa+i \omega)/(\kappa-i \omega)$, and $\sigma_3(\omega)=-r$. When $ S_{d}^{X}(\omega)=1/2$ it is convenient to rewrite in a new form both the Heisenberg inequality of Eq.~(\ref{heisen}) and $ S_{d}^{opt}(\omega)$. In fact, the Heisenberg inequality becomes $S_{d}^{Y}(\omega) \geq 2 [S_{d}^{XY}(\omega)]^2+1/2$, which suggests the following parametrization
\begin{equation}\label{paraspey}
    S_{d}^{Y}(\omega) = 2 [S_{d}^{XY}(\omega)]^2+\frac{1}{2}+S_{d}^{r}(\omega),
\end{equation}
so that the Heisenberg condition becomes simply $S_{d}^{r}(\omega)\geq 0$ with $S_{d}^{r}(\omega)$ a ``residual'' spectrum measuring the distance from the minimum uncertainty condition. In the resonant case the optimal squeezing spectrum can be rewritten as
\begin{equation}\label{speoptsr}
   S_d^{opt}(\omega)=\frac{S_{d}^{r}(\omega)+1/2}{1+S_{d}^{r}(\omega)+2 [S_{d}^{XY}(\omega)]^2+\sqrt{\left[S_d^{r}(\omega)
    +2 [S_{d}^{XY}(\omega)]^2\right]^2+
    4\left[S_d^{XY}(\omega)\right]^2}},
\end{equation}
showing two important aspects of ponderomotive squeezing at resonance. First of all one has that \emph{the optimal squeezing spectrum is squeezed as soon as $S_d^{XY}(\omega) \neq 0$}. However this result is not relevant by itself in practice because, as mentioned in the Introduction, the present treatment neglects technical noise contributions, such as electronic noise and phase and amplitude noise of the driving laser. This technical noise adds to the expected noise spectrum and, if $ S_d^{opt}(\omega)$ is not too much below the shot-noise limit, it prevents the observation of squeezing. Therefore, one has to make $S_d^{opt}(\omega)$ as small as possible, and Eq.~(\ref{speoptsr}) suggests how this can be achieved. In fact, the strongest squeezing is obtained when the two limits $S_{d}^{r}(\omega) \ll 1$ and $\left[S_d^{XY}(\omega)\right]^2 \gg 1$ are simultaneously satisfied:
\begin{equation}\label{speoptsrlim}
   S_d^{opt}(\omega) \simeq \frac{1}{8 \left[S_d^{XY}(\omega)\right]^2} \;\; {\rm for}\; S_{d}^{r}(\omega) \ll 1,\;\;\left[S_d^{XY}(\omega)\right]^2 \gg 1.
\end{equation}
This strong squeezing is obtained at the optimal phase
\begin{equation}\label{phioptlim}
   \phi_{opt}(\omega)\simeq -\frac{1}{2}\arctan\left[\frac{1}{S_d^{XY}(\omega)}\right],
\end{equation}
which is very close, but strictly different from $\phi=0$, where the corresponding quadrature $X_d$ is just at the shot-noise limit (see Eq.~(\ref{speampli})). This means that at fixed frequency, squeezing is achieved only within a narrow interval for the homodyne phase around $\phi_{opt}(\omega)$, of width
$$
\delta \phi^{sq}(\omega) \sim 2 \left |\phi_{opt}(\omega)\right| \sim \arctan\left|\frac{1}{ S_d^{XY}(\omega)}\right|.
$$
This extreme phase dependence is a general and well-known property of quantum squeezing, which is ultimately due to the Heisenberg principle: the width of the interval of quadrature phases with noise below the shot-noise limit is inversely proportional to the amount of achievable squeezing. This implies that in order to detect squeezing one has to tune and stabilize the phase of the homodyne detection apparatus with extreme accuracy.

By using the general equations of the former section, one gets the explicit expressions for $S_d^{XY}(\omega)$ and $S_d^{r}(\omega)$ at resonance as a function of the parameters of the optomechanical system:
\begin{eqnarray}\label{spexyespl}
   && S_d^{XY}(\omega)=\kappa t^2 {\rm Re}\left\{\frac{\lambda_G(\omega)}{(\kappa+i\omega)\left[\kappa-i \omega+\sqrt{\eta}r\sin \theta \lambda_g(\omega)\right]}\right\},\\
    && S_d^{r}(\omega)=\frac{2r^2}{t^2}[S_d^{XY}(\omega)]^2 + \frac{2\kappa t^2 {\rm Im}\left\{\lambda_G(\omega)\right\}\coth\left(\hbar \omega/2\kappa_B T\right) }{\left|\kappa-i
    \omega+r\sqrt{\eta}\sin \theta \lambda_g(\omega)\right|^2} \nonumber \\
   && + 2\kappa^2 t^2{\rm Im}\left\{\frac{\lambda_G(\omega)}{(\kappa+i\omega)\left[\kappa-i \omega+\sqrt{\eta}r\sin \theta \lambda_g(\omega)\right]}\right\}^2\nonumber \\
    && +\frac{t^2}{2}\frac{|\lambda_g(\omega)|^2 (\kappa^2+\omega^2)-4 \kappa r\sqrt{\eta}\cos\theta {\rm Re}\left\{\lambda_G^*(\omega)\lambda_g(\omega)(\kappa+i\omega)\right\}}{(\kappa^2+\omega^2)\left|\kappa-i
    \omega+r\sqrt{\eta}\sin \theta \lambda_g(\omega)\right|^2} \label{sperespl}.
\end{eqnarray}
These two equations show how to choose the system parameters in order to optimize ponderomotive squeezing. The fact that $S_d^{r}(\omega)$ must be as small as possible while keeping $S_d^{XY}(\omega)$ very large implies first of all that one has to take $t  \to 1$, i.e., the output beam splitter must possess high transmissivity, so that the first contribution of Eq.~(\ref{sperespl}) to $S_d^{r}(\omega)$ is negligible. This means that only a tiny fraction of the cavity output light is used in the feedback loop, but this can always be compensated by adjusting the feedback gain, i.e., the modulus of $\lambda_g(\omega)$. The second term of Eq.~(\ref{sperespl}) is the thermal noise contribution to the spectrum, which is smaller at low temperatures. However, apart from lowering temperatures, the most efficient way to suppress the thermal noise contribution is to choose a mechanical resonator with normal modes possessing large quality factors $Q_j=\omega_j/\gamma_j$. In fact
\begin{equation}\label{img}
    {\rm Im}\left\{\lambda_G(\omega)\right\}=\sum_j \frac{G_j^2\omega_j \gamma_j \omega}{(\omega_j^2-\omega^2)^2+\omega^2\gamma_j^2},
\end{equation}
so that
\begin{eqnarray}
  {\rm Im}\left\{\lambda_G(\omega)\right\} &\simeq & \omega \sum_j \frac{G_j^2}{\omega_j^2 Q_j}, \;\;\;\omega \ll \omega_j,\\
 {\rm Im}\left\{\lambda_G(\omega)\right\}&=& \sum_j \frac{G_j^2 Q_j}{\omega_j}, \;\;\;\omega = \omega_j.
\end{eqnarray}
This means that choosing large values of $Q_j$ has the effect of concentrating the thermal noise contribution only within the frequency bands corresponding to the very narrow (width $\sim \omega_j/Q_j$) mechanical resonance peaks, while the value of ${\rm Im}\left\{\lambda_G(\omega)\right\}$ is negligible in a wide low-frequency bandwidth below the mechanical resonances. Also the third contribution of Eq.~(\ref{sperespl}) becomes very small when ${\rm Im}\left\{\lambda_G(\omega)\right\}$ is small and therefore a high-quality mechanical resonator is able to suppress also this contribution to the output homodyne spectrum. Therefore we expect that the optimal squeezing spectrum is well below the shot-noise limit in a wide low-frequency band below the mechanical resonances, while being at shot-noise level at mechanical resonance frequencies.

The fourth contribution to $S_d^{r}(\omega)$ is related to the optical input vacuum noise injected in the system by the feedback loop, and it describes the main effect of the feedback control on the squeezing spectrum. This fourth term may become negative, showing that \emph{feedback may enforce squeezing}, i.e., one may have feedback-assisted ponderomotive squeezing. Considering frequencies $\omega \ll \kappa$, exploiting that $\lambda_G(\omega)$ is essentially real over a wide frequency range, and that $r \ll 1$, one has that the feedback-induced contribution to the spectrum is minimum (and negative) when \begin{equation}\label{optfeed0}
\lambda_g(\omega)\simeq 2 r\sqrt{\eta}\cos\theta \lambda_G(\omega),
\end{equation}
which can be satisfied for different choices of the feedback transfer functions $ g_j(\omega)$ in general. If we look for a solution valid at all frequencies $\omega$, the simplest choice is to take
\begin{equation} \label{optfeed}
 g_j(\omega) \simeq 2 r\sqrt{\eta}\cos\theta G_j \;\;\; \forall j,
 \end{equation}
that is a \emph{frequency-independent proportional} control, which is causal and very different from the derivative (viscous) control employed in cold damping schemes for cooling the vibrational modes \cite{cohadon99,bouwm,rugar,vinante,bowen10}. Eq.~(\ref{optfeed}) defines the optimal feedback control for generating ponderomotive squeezing, which is however difficult to implement because it requires $g_j \propto G_j$, $\forall j$. In fact, this means a feedback transducer which couples with each vibrational mode of the resonator in the same way as the cavity mode does. This situation could be, at least approximately, realized by applying the feedback control on the mechanical element through the radiation pressure of an additional laser beam with the same spatial transverse profile of the cavity field. However, the general condition of Eq.~(\ref{optfeed0}) is easier to satisfy, without requiring the condition of Eq.~(\ref{optfeed}), if one wants to improve ponderomotive squeezing only within a not too broad frequency interval. For example, at low frequencies, below all mechanical resonances $\omega \ll \omega_j$, the unperturbed susceptibilities $\chi_j^{(0)}(\omega)$ of Eq.~(\ref{susczero}) and therefore $\lambda_G(\omega)$ are frequency-independent, so that Eq.~(\ref{optfeed0}) is satisfied by taking again a proportional control and simply adjusting the overall gain of the feedback loop.

The noise subtraction caused by the optimal feedback of Eq.~(\ref{optfeed}) is
\begin{equation}\label{subtra}
\Delta S_d^r \sim -4 r^2 \eta \cos^2\theta  \frac{\lambda_G^2}{\kappa^2} \sim -4 r^2 \eta \cos^2\theta \left[ S_d^{XY}\right]^2,
\end{equation}
which however can never be too negative; in fact, under optimal conditions, even though $S_d^{XY} \to \infty$, one has $r^2  \left[ S_d^{XY}\right]^2 \to 0$, which stems from the fact that $S_{d}^{r}(\omega)$ must approach zero (see the first term in Eq.~(\ref{sperespl})). This feedback-induced noise subtraction is due to destructive interference between the intracavity field fluctuations and the fluctuations injected by the feedback loop and transferred into the cavity by the optomechanical coupling.

Feedback also modifies the cavity response as described by the expression of $D(\omega)$ of Eq.~(\ref{denom}), and this modification of the cavity and mechanical resonator response is just at the basis of the feedback cooling process proposed in \cite{Mancini98,courty,vitalirapcomm,quiescence02,genes07} and experimentally demonstrated in \cite{cohadon99,arcizet06b,bouwm,rugar,vinante,bowen10}.
On the contrary, we have seen that feedback modification of cavity response is not useful for the generation of squeezing and its only relevant effect is of making the system more easily unstable (see also Eq.~(\ref{stabres})). This means that the feedback optimal for squeezing corresponds to $\theta =0$, because in this case there is no feedback-induced modification of the cavity, the system is always stable, and at the same time the feedback-induced noise subtraction of Eq.~(\ref{subtra}) is maximum. Choosing $\theta =0$ means measuring the \emph{amplitude} quadrature of the mode reflected at the output beam splitter, which at resonance, is not coupled to the mechanical element, and therefore in this case the improvement of squeezing due to feedback and described by Eq.~(\ref{subtra})
is not determined by the feedback modification of the cavity dynamics but only by the destructive interference between the intracavity fluctuations and those injected by the feedback loop.

The performance of such optimized feedback control ($t=0.99$, $\theta=0$, $\eta=1$) is shown in Fig.~2, where the optimal squeezing spectrum $S_d^{opt}(\omega)$ is shown (full line) and compared to the case without feedback (dashed line). The plot refers to a cavity with bandwidth $\kappa \simeq 1$ MHz, corresponding to a finesse $F \simeq 8000$, length $L=6$ cm, driven by a laser at $1064$ nm and with input power $\mathcal{P}_{in} =30$ mW. We have considered a mechanical resonator with a number of vibrational modes with resonance frequencies between $150$ and $600$ KHz, all with the same effective mass $m_j=100$ ng, and with the same quality factor $Q_j=10^4$, placed at a temperature $T=4$ K. We see that the optimal squeezing spectrum is well below the shot noise limit out of the mechanical resonance peaks, both with and without feedback, due to the fact that with the chosen parameter values, the radiation pressure interaction predominates over thermal noise. Feedback control provides a visible, even if not macroscopic, noise reduction over a wide frequency range. In Fig.~2b the same curves in the low-frequency band below the resonance peaks are shown, and one sees that the noise spectrum is roughly three times smaller due to feedback.

\begin{figure}[tb]
\centerline{\includegraphics[width=0.99\textwidth]{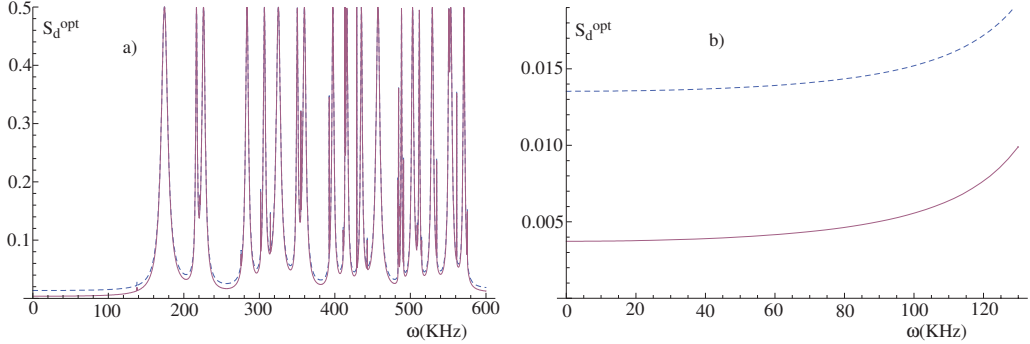}} \caption{Optimal spectrum of squeezing in the presence (full line) and in the absence of feedback (dashed line). Feedback improves squeezing in a visible way. We have considered a cavity with bandwidth $\kappa = 1$ MHz, length $L=6$ cm, driven by a laser at $1064$ nm and with input power $\mathcal{P}_{in} =30$ mW. The mechanical resonator has $25$ vibrational modes with resonance frequencies between $150$ and $600$ KHz, all with the same effective mass $m_j=100$ ng, and with the same quality factor $Q_j=10^4$, placed at a temperature $T=4$ K. We have also taken optimal conditions for the feedback control: $t=0.99$, $\eta =1$, $\theta =0$, $g_j\simeq 2r\sqrt{\eta}\cos\theta G_j = 0.28 G_j$. Fig.~2b is a zoom of the low frequency region below 120 KHz.}\label{fig2}
\end{figure}

In Fig.~3a we show the optimal field phase $\phi_{opt}(\omega)$ corresponding to the optimal squeezing spectrum shown in Fig.~2. The curves with and without feedback are indistinguishable and in both cases the optimal phase is practically frequency-independent in the low-frequency band up to $100$ KHz. Fig.~3b shows the noise spectrum of a \emph{given} quadrature with phase $\phi$ (at the fixed frequency $\omega=10$ KHz), versus $\phi$, for the same parameters of Fig.~2. At the small, nonzero value corresponding to $\phi_{opt}(\omega)$ of Fig.~3a, one gets up to $20$ dB of ponderomotive squeezing, but only within an extremely narrow phase interval around $\phi_{opt}(\omega)$, as it always occurs for squeezing, since the width of the useful phase interval is inversely proportional to the maximum achievable squeezing.

\begin{figure}[tb]
\centerline{\includegraphics[width=0.99\textwidth]{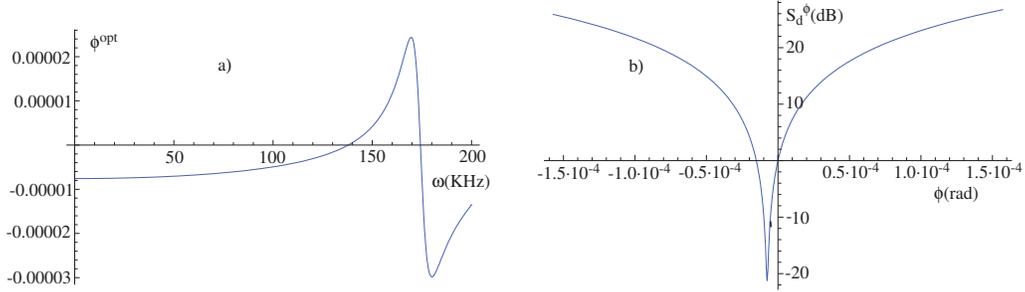}} \caption{a) Optimal field phase $\phi_{opt}(\omega)$ corresponding to the optimal squeezing spectrum shown in Fig.~2. The curves in the case with and without feedback are indistinguishable. b) Noise spectrum (in dB) of the quadrature at phase $\phi$ at the fixed frequency $\omega=10$ KHz, $S_d^{\phi}$, versus the phase $\phi$, with (full line) and without feedback (dashed line). Parameters are the same as in Fig.~2.}\label{fig3}
\end{figure}

\section{Conclusions}

We have presented a general description of ponderomotive squeezing and discussed if and how an added feedback control may improve squeezing at the cavity output. Appreciable ponderomotive squeezing is achieved in general when the noise correlations caused by radiation pressure predominates over thermal noise, i.e., when $2[S_d^{XY}(\omega)]^2 \gg S_d^r(\omega)$ (see Sec.~V). Using the above results this condition can be written as
\begin{equation}
\frac{2[S_d^{XY}(\omega)]^2}{S_d^r(\omega)} \sim \frac{\left[{\rm Re}\{\lambda_G(\omega)\}\right]^2}{\kappa \coth(\hbar \omega/2 k_B T){\rm Im}\{\lambda_G(\omega)\}} \sim \frac{\mathcal{P}_{in} \omega_0}{m c^2 \omega_m^2}\frac{\mathcal{F}^2 Q}{\bar{n}} \gg 1,
\end{equation}
where in the last ratio we have assumed $\omega \sim 0$, considered for simplicity a single vibrational mode with mass $m$, quality factor $Q$, frequency $\omega_m$, and we have denoted with $\bar{n}=k_B T/\hbar \omega_m$ its mean thermal vibrational number. $\mathcal{F}=\pi c/2\kappa L$ is the cavity finesse.

We have seen that by adding a suitable feedback control one can get a moderate improvement of ponderomotive squeezing thanks to a destructive interference between the intracavity light fluctuations and those injected by the feedback loop. The optimal feedback control is obtained by employing an highly transmitting output beam splitter ($r \to 0$), by measuring the amplitude quadrature ($\theta =0$), and by adopting a proportional control such that $g_j(\omega) \propto G_j$.

\section{Acknowledgements}

This work has been supported by the European Commission (FP-7 FET-Open project MINOS), and by INFN (SQUALO project).



%% References


\bibliographystyle{elsarticle-num}
\bibliography{<your-bib-database>}



\end{document}